# Nanoscale Voltage Enhancement at Cathode Interfaces in Li-ion Batteries

Shenzhen Xu[1], Ryan Jacobs[2], Chris Wolverton[4], Thomas Kuech[3] and Dane Morgan[1,2]

[1]Materials Science Program, University of Wisconsin- Madison, Madison, WI, USA

[2]Department of Materials Science and Engineering
University of Wisconsin – Madison, Madison, WI, USA

[3]Department of Chemical and Biological Engineering
University of Wisconsin – Madison, Madison, WI, USA

[4]Department of Materials Science and Engineering
Northwestern University, Evanston, IL, USA

## Abstract:

Interfaces are ubiquitous in Li-ion battery electrodes, occurring across compositional gradients, regions of multiphase intergrowths, and between electrodes and solid electrolyte interphases or protective coatings. However, the impact of these interfaces on Li energetics remains largely unknown. In this work, we calculated Li intercalation-site energetics across cathode interfaces and demonstrated the physics governing these energetics on both sides of the interface. We studied the olivine/olivine-structured $Li_xFePO_4/Li_xMPO_4$ (x=0 and 1, M=Co, Ti, Mn) and layered/layered-structured $LiNiO_2/TiO_2$ interfaces to explore different material structures and transition metal elements. We found that across an interface from a high- to low-voltage material the Li voltage remains constant in the high-voltage material and decays approximately linearly in the low-voltage region, approaching the Li voltage of the low-voltage material. This effect ranges from 0.5-9nm depending on the interfacial dipole screening. This effect provides a mechanism for a high-voltage material at an interface to significantly enhance the Li intercalation voltage in a low-voltage material over nanometer scale. We showed that this voltage enhancement is governed by a combination of electron transfer (from low- to high-voltage regions), strain and interfacial dipole screening. We explored the implications of this voltage enhancement for a novel heterostructured-cathode design and redox pseudocapacitors.

# 1 Introduction:

Many methods have been developed to improve the performance of battery electrodes over the past 25 years, particularly cathodes, and in many modern commercial systems the cathode is a complex, heterogeneous material. In these highly engineered materials, interfaces are ubiquitous and can have a significant influence on the behavior of intercalating ions. For example, material intergrowths with nanoscale regions of multiple phases in close contact, such as γ-$MnO_2$[1,2] and Li-excess materials[3,4], are some of the most widely used commercial and most promising research materials in primary alkaline and Li-ion batteries, respectively. The exceptional performance of these materials appears to be directly coupled to the presence of the nanoscale regions and their interfaces, as this performance is greatly reduced in the bulk phases of the components.

In Li-ion batteries, interfaces also occur in compositionally-graded cathodes such as those developed for layered $Li[Ni_{1-x}M_x]O_2$ compounds, which contain a central bulk rich in Ni and an outer layer rich in Mn and Co[5]. The concentration gradient in $Li[Ni_{1-x}M_x]O_2$ creates multiple, diffuse interfaces between phases which contain different transition metal concentrations. Similar concentration gradient-induced interfaces occur in the spinel-structured $Li_yVa_{1-y}Mn_xNi_{2-x}O_4$ (Va= vacancy) cathode material with a compositional core-shell structure[6]. Moreover, interfaces inevitably occur for cathode particles at their surface, either with a solid electrolyte interphase (SEI) or frequently a protective coating layer, the latter of which is typically comprised of insulating metal oxides or fluorides[7-15] or sometimes an electrochemically active material (e.g., Li-Ni-$PO_4$ coating layer)[16].

Despite the importance of interfaces for electrode properties in batteries, there is still little knowledge specifically related to the Li intercalation voltage in materials near cathode interfaces and the mechanisms which control this voltage. In particular, materials with different bulk Li intercalation voltages (e.g. in compositionally-graded cathodes, cathode/coating interfaces, etc.) are expected to influence each other across an interface, but neither the nature of that influence nor over what range it might persist is known. Previous works on interfaces have provided insights into some key mechanisms that play a significant role in Li intercalation energetics near the interface. For example, Liu et al. [17] studied a pseudocapacitive effect on interfacial Li storage in the graphene/$TiO_2$ (anatase) interface, demonstrating that the electron transfer to the graphene layer and the existence of interfacial oxygen atoms contribute to extra Li storage in the graphene/$TiO_2$ interface. Haruyama et al. [18] worked on the interfacial space charge layer effect on the interface between oxide cathode material and solid-state sulfide electrolyte and demonstrated that the interfacial dipole (caused by the space charge layer) and the local microstructure at the interface has a significant impact on Li distribution near the interface. These previous works indicate that the electron transfer (from intercalated Li), interfacial dipoles and the local structure may affect the Li energetics near the electrode interface. However, it is still unclear what the length scale of these interfacial effects is and how these physical mechanisms (e.g., electron transfer, interfacial dipole) explicitly change the Li energetics (relative to the bulk case) at different Li intercalation sites near the interface.

In this work, we explored the mechanisms and the intercalation site position dependence of interfacial effects on Li energetics and intercalation. We focused on interfaces between two cathode materials with significantly different bulk Li intercalation voltages. We studied both the olivine/olivine-structured $Li_xFePO_4$/$Li_xMPO_4$ (x=0 or 1, M=Co, Ti, Mn) and layered/layered-structured $LiNiO_2$/$TiO_2$ interfaces to explore different material structures and a range of transition metal elements. The use of coherent olivine-structured and layered-structured material interfaces mitigates the complicated issues of modeling the generally unknown structure of the interface between dissimilar crystal structures. We calculated Li intercalation site energy profiles across the interfaces and studied the physics governing the Li intercalation site energy behavior at different positions of both sides of the interfaces. It is expected the phenomena and the governing physics investigated here to be applicable to general materials systems, and the results suggest new approaches for the design of novel electrode superlattice structures for future applications in Li-ion batteries and redox pseudocapacitors.

The main results contained in this work are outlined as follows. In **Section 2.1**, we talked about the nature of the interface we studied and the ideality of our modeling system. In **Section 2.2,** we calculated the Li intercalation site energy profile as a function of distance from the interface for the olivine/olivine-structured $Li_xFePO_4/Li_xMPO_4$ (x=0 or 1, M=Co, Ti, Mn) interface. This energy profile shows a nanoscale voltage enhancement effect in the low-voltage region, and we have proposed a charge-transfer mechanism to explain the physics governing the behavior of the Li energetics. In **Section 2.3** we examined the screening effect of the intercalated Li atoms in the olivine/olivine interface, which is manifested as a modification of the interfacial dipole from charge transfer. In **Section 2.4**, we introduced the layered/layered-structured $LiNiO_2/TiO_2$ interface and demonstrated how a multiple-redox transition metal cation ($Ni^{2+\rightarrow 4+}$) can be utilized by an unscreened Li through an interface over ~9 nm. We also demonstrated that the screening effects resulting from layer-by-layer Li intercalation effectively restricts this interfacial effect to a region of 2-3 atomic layers at the interface. In **Section 3**, we proposed a possible cathode heterostructure design of alternating low- and high- voltage materials that takes advantage of the nanoscale voltage enhancement effect. We also discussed implications of our single Li intercalation results (shown in **Section 2.2** and **2.4**) for redox pseudocapacitors[19].

## 2 Results and Discussion:

### 2.1 Nature of the interfaces studied

It should be noted that the calculations in this paper are based on ideal, coherent interfaces. Due to its grounding in basic energetics of redox, dipoles, and electrostatics, we expect the physics identified and discussed in this work to occur in more realistic interfaces as well, although the magnitudes of the contributing effects may be impacted by the specific interfacial structure. In particular, realistic interfaces or heterostructures may not have the sharp termination of the chemical composition modeled in our calculations. The different transition metal elements in the two sides of the interface may exchange across the interface to form interlayers. This chemical-composition mixing phenomenon can easily happen during the synthesis process or heating, and may even occur during electrochemical cycling at room temperature, depending on the mobility of the cations. Based on the understanding of the origins of interfacial phenomena developed in this work, we believe that the voltage enhancement effect discussed here should still hold when there is transition metal cation mixing at the interface. We also note that the layered-structured $TiO_2$ modeled in our $LiNiO_2$-$TiO_2$ system in Section 2.4 is not a known polymorph of $TiO_2$ and may be unstable or metastable if made experimentally. However, the structure is stable in the calculations and serves as illustrative example of a high-voltage and low-voltage interface of layered materials which is expected to apply to layered and other structures in general.

## 2.2 Li intercalation site energetics across interfaces

The Li intercalation site voltage profiles across the olivine/olivine-structured $Li_xFePO_4$/$Li_xMPO_4$ (x=0 or 1, M=Co, Ti, Mn) interfaces are shown in **Figure 1**. Details regarding the definition and calculation methods of Li intercalation site voltage can be found in the Computational Methods in **Section 5**. We have calculated the case of single Li intercalation in a fully-delithiated olivine interface cell $Li_xFePO_4$/$Li_xMPO_4$, x=0 (this fully-delithiated limit corresponds to the beginning of the discharging process) and the case of single Li deintercalation out of a fully-lithiated olivine cell $Li_xFePO_4$/$Li_xMPO_4$, x=1 (this fully-lithiated limit corresponds to the beginning of the charging process). It should be noted here that the results of this section are an exploration of Li intercalation/deintercalation site energetics at different positions near the interface. The real discharging/charging order of Li intercalation/deintercalation is not considered in this section, and it is instead discussed in the **Section 2.3 and 2.4**. Here the "real discharging/charging order" means the order in which Li enter and leave the material during a real intercalation/deintercalation process, which follows thermodynamic rules of minimizing free energy, i.e., Li atoms go into/out of the cathode material one by one in the intercalation/deintercalation order of high→low/low→high voltage sites. This definition is what is meant by "real discharging/charging order" phrases in the following text.

In all three interfaces $Li_xFePO_4/Li_xMPO_4$ (x=0, M=Co, Mn, Ti) in the fully-delithiated limit, the Li intercalation voltage profile has a plateau in the material with the higher intercalation voltage. When the Li intercalates in the low-voltage material near the interface, there is an approximately linear voltage drop away from the interface into the low-voltage region. The voltage profile again reaches a plateau in the lower voltage material when the Li intercalation site is far enough away from the interface (≈1-2 nm). The voltage values in the plateau regions are equal to the bulk Li intercalation voltages for each corresponding material. Thus, at the approximate distance of 1-2 nm away from the interface, the interfacial effects become negligible. It is clear that the intercalation voltage in the low-voltage material is significantly enhanced relative to its bulk value within a ≈1-2 nm distance from the interface. The deintercalation voltage behavior of the fully-lithiated limit case is analogous to the intercalation voltage behavior of the delithiated case. The only difference between deintercalation compared to intercalation is that for deintercalation the voltage is a plateau in the low-voltage region and gradually increases and approaches the high-voltage plateau in the high-voltage material. The length scale of the interfacial effect is also about ≈1-2 nm in the fully-lithiated limit case. We will use the fully-delithiated limit (one Li intercalation in a fully-delithiated interface cell) as the representative case for the exploration of the underlying physics in **Section 2.2.2**. As we will show in **Section 2.2.2**, the governing physics of the two limiting cases is the same.

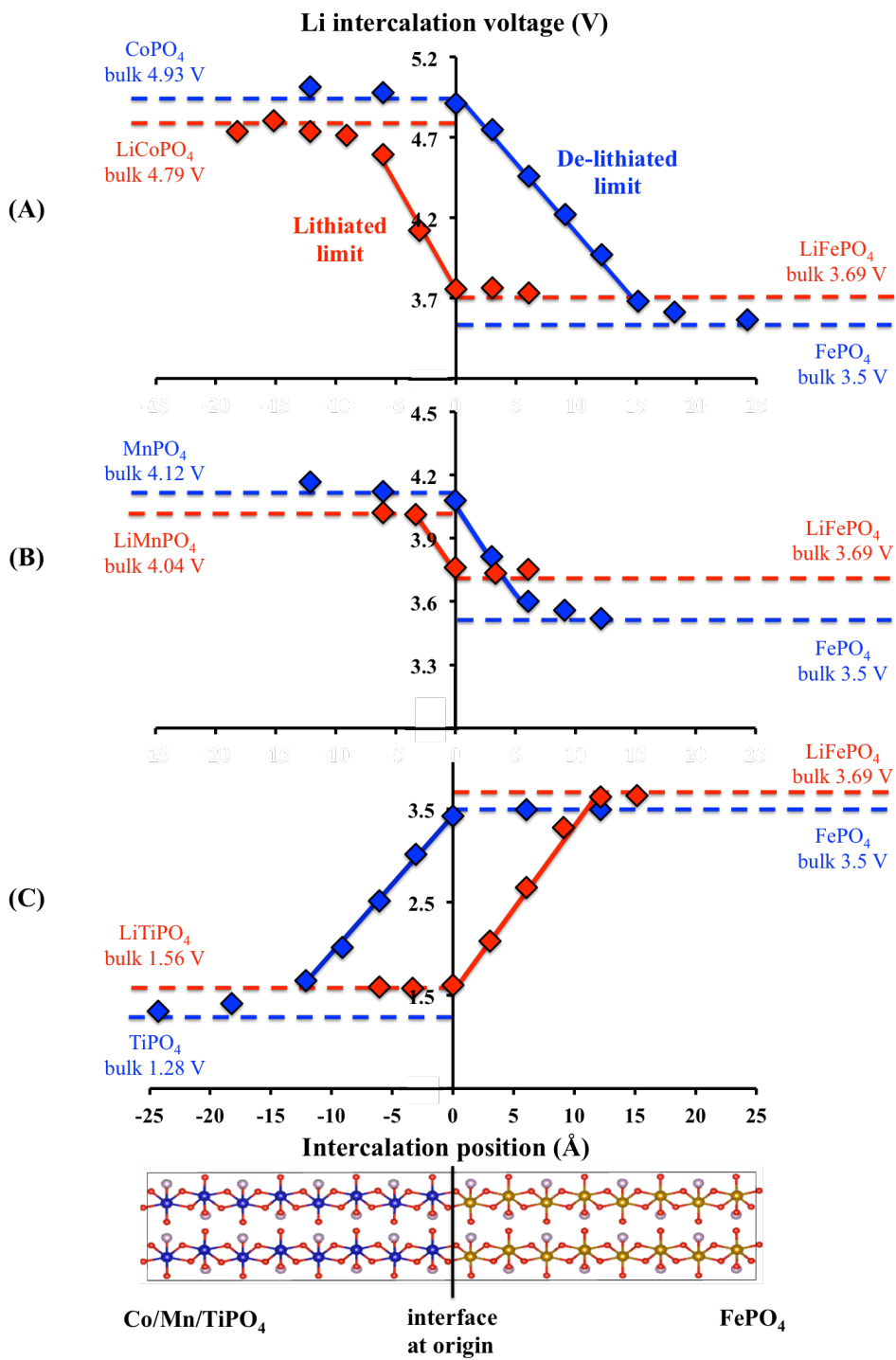

**Figure 1.** Li intercalation voltage profiles across the interfaces of (a) $Li_xFePO_4/Li_xCoPO_4$; (b) $Li_xFePO_4/Li_xMnPO_4$; (c) $Li_xFePO_4/Li_xTiPO_4$ in the delithiated limit x=0 (Li intercalation in $FePO_4/MPO_4$, blue lines) and the lithiated limit x=1 (Li deintercalation in $LiFePO_4/LiMPO_4$, red lines). The atomic structure of the olivine material interface is shown at the bottom. The interface is at the origin where the intercalation position is referred to as 0 in the figure. In (a)-(c), the dashed lines indicate the bulk voltages far from the interface, and the solid lines serve as guides

for the eye in the region of approximately linear change in voltage with distance from the interface.

### 2.2.1 Terms controlling energetics across the interface

We propose that we can write the Li site intercalation energy (and thus site voltage) relative to the bulk material far from the interface, $\Delta E_{\text{intercal}}(x_{Li})$, in terms of four contributions, all of which are a function of the Li intercalation position $x_{Li}$. These contributions are summarized as:

$$\Delta E_{\text{intercal}}(x_{Li}) = \Delta E_{\text{distortion}}(x_{Li}) + \Delta E_{\text{e-redox}}(x_{Li}) + \Delta E_{\text{e-dipole}}(x_{Li}) + \Delta E_{\text{e-chg.sep.}}(x_{Li}). \qquad (1)$$

$\Delta E_{\text{intercal}}(x_{Li})$ is the change of the Li intercalation energy when Li is intercalated at $x_{Li}$ from the interface (numerator of **Equation 2** in **Section 5**) relative to the bulk intercalation energy of the high-voltage material (note that the reference state can be bulk intercalation energy in either the high-voltage material or low-voltage material, and we use the high-voltage material here for the following discussion).

The first contribution, $\Delta E_{\text{distortion}}(x_{Li})$, is the lattice distortion energy. The lattice distortion energy depends on the mechanical response of the material, which couples to the local lattice distortion caused by the small expansion of the crystal lattice due to the Li insertion. $\Delta E_{\text{distortion}}(x_{Li})$ is the relative change of lattice distortion energy caused by the Li intercalation at $x_{Li}$ to the reference lattice distortion energy of bulk Li intercalation in the high-voltage material. Since this is a relative lattice distortion energy, it will have a small value when the low- and high-voltage materials are similar in structure.

The second contribution, $\Delta E_{\text{e-redox}}(x_{Li})$, is the transition metal redox energy change relative to the reference transition metal redox energy in the bulk high-voltage material. The redox energy is the energy change resulting from modification of the transition metal oxidation state due to the intercalated Li valence electron. $\Delta E_{\text{e-dipole}}(x_{Li})$ is the energy change associated with moving the Li valence electron across the interfacial dipole when charge transfer across the interface occurs. Finally, $\Delta E_{\text{e-chg.sep.}}(x_{Li})$ is the electrostatic charge-separation energy of the Li valence electron interacting with Li$^+$ due to the charge transfer across the interface. This energy only takes account of the electrostatic interaction between the valence electron (redistributed to the high-voltage material) and the $Li^+$, which is a function of the distance the valence electron is moved from the intercalated $Li^+$, and acts to increase the intercalation energy and therefore decrease the voltage.

$\Delta E_{\text{e-dipole}}(x_{Li})$ is perhaps more physically complex than the other terms so we explain it in more detail here. In the region where charge transfer (from the low- to high-voltage region) occurs, there will be a modification of the interfacial dipole caused by the intercalated Li. The reference dipole value is the initial interfacial dipole $P_0$ created as a result of interfacial chemical bonding between dissimilar materials, prior to any Li intercalation. When Li atoms are intercalated in the low-voltage material near the interface, the interfacial dipole is modified and its value is changed to $P_{\text{lithiated}}$ due to the charge transfer from the Li to the high-voltage material ($P_{\text{lithiated}}$ depends on the amount of Li intercalated in the low-voltage region near the interface as well as $x_{Li}$). $\Delta E_{\text{e-dipole}}(x_{Li})$ in **Equation 1** is the energy change in the Li intercalation associated with moving the Li valence electron across the dipole difference between the modified interfacial dipole and the initial interfacial dipole ($\Delta P = P_{\text{lithiated}} - P_0$).

Overall, the above four key terms (Eq. (1)) that determine the Li intercalation site energy profile can be sorted into an electronic part, an ionic part and the electron-ion interaction part. The electronic part contains the redox energy of the electron reducing a certain transition metal $\Delta E_{\text{e-redox}}$ and the energy of the electron moving through an interfacial dipole $\Delta E_{\text{e-dipole}}$, The ionic part corresponds to the lattice distortion energy $\Delta E_{\text{distortion}}$ due to the insertion of an ion into the lattice, and the electron-ion interaction part is the electrostatic energy $\Delta E_{\text{e-chg.sep.}}$ of separating the negatively-charged electron from the positively-charged ion.

### 2.2.2 Interfacial effects in the olivine-structured interface

Now we explore how each of these terms is realized in the $Li_xFePO_4/Li_xMPO_4$ (x=0, M=Co, Mn, Ti) systems, using the $CoPO_4/FePO_4$ interface as our main example. In the fully-delithiated limit, when the Li intercalation position is in the low-voltage material within a ≈1-2 nm distance from the interface, charge transfer from the Li across the interface to the high-voltage region occurs. The schematic plot to explain this charge transfer mechanism is shown in **Figure 2** for the CoPO$_4$/FePO$_4$ interface. In general, electrons will preferentially move from a low-voltage region to a high-voltage region. As shown in **Figure 2**, the Li valence electron is generally expected to localize on one or both of the transition metal species, or remain affixed to Li to maintain a neutral Li atom.

As the crystal structures of all olivine materials considered here (Fe/Co/Mn/TiPO$_4$) are the same and the lattice parameter mismatch between them is small (<2% for all the x, y, z directions, discussed in the **Supplemental Information (SI) Section 1.1**), $\Delta E_{\text{distortion}}(x_{Li})$ is approximately zero regardless of the Li intercalation position. However, $\Delta E_{\text{e-redox}}$ depends on $x_{Li}$ and is a function of the transition metal species being reduced. Based on our definition of $\Delta E_{\text{e-redox}}$ where the high-voltage material is taken as the reference material, $\Delta E_{\text{e-redox}} = 0$ when a Li atom is intercalated in a site in the high-voltage material and reduces the high-voltage transition metal. When the Li is in a site in the low-voltage material far from the interface (low-voltage plateau region in **Figure 1**), thus reducing the transition metals in the low-voltage material, $\Delta E_{\text{e-redox}}$ is equal to the redox energy difference of the low- minus the high-voltage transition metal elements. For example, in the case of FePO$_4$/CoPO$_4$, the bulk Li intercalation voltage of FePO$_4$ is about 1.4 V lower than that of CoPO$_4$, which means the redox energy of Fe is approximately 1.4 eV higher than that of Co, and $\Delta E_{\text{e-redox}} = 1.4$ eV for a Li in a site in FePO$_4$. If Li is intercalated in the low-voltage material near the interface, it is energetically favorable for the Li valence electron to transfer across the interface to reduce the high-voltage transition metal, as shown in **Figure 2**. Therefore, in the voltage enhanced region of the low-voltage material, $\Delta E_{\text{e-redox}}$ is also equal to zero, just as in the high-voltage material.

In the case of a single Li intercalation, the dipole that the Li valence electron moves across is close to $P_0$. Therefore, in this case, $\Delta E_{\text{e-dipole}}(x_{Li})$ is approximately equal to zero. In the following discussion (**Section 2.3**) where the intercalation of multiple Li is considered, the $\Delta E_{\text{e-dipole}}(x_{Li})$ term will no longer be zero. In this case, the interfacial dipole enhancement $\Delta P$ caused by the intercalated Li in the low-voltage material near the interface will result in a pronounced screening effect.

The final term to consider is $\Delta E_{\text{e-chg.sep.}}$. In the intercalation voltage results of FePO$_4$/CoPO$_4$ we find that when Li is in FePO$_4$ near the interface, although the Li valence electron localizes on the Co atoms by moving across the interface, the intercalation voltage is reduced relative to the bulk value of $CoPO_4$. This reduction is primarily due to the electrostatic energy cost of separating the electron from the Li, which is captured in the term $\Delta E_{\text{e-chg.sep.}}$.

If we consider the four energy terms in **Equation 1** together, the qualitative explanation for the voltage profile across the interface between a high- and low-voltage material for the case of single Li intercalation (explained here for the example of $FePO_4/CoPO_4$) can be summarized as follows: a) when the intercalation site is in the $CoPO_4$ region, all four terms are equal to zero, so $\Delta E_{\text{intercal}}(x_{Li})$ is zero and thus the Li intercalation voltage is equal to the bulk value of $CoPO_4$; b) when the intercalation site is in $FePO_4$ near the interface, $\Delta E_{\text{distortion}}(x_{Li})$, $\Delta E_{\text{e-redox}}(x_{Li})$ and $\Delta E_{\text{e-dipole}}(x_{Li})$ are approximately zero, and the electrostatic charge-separation energy term $\Delta E_{\text{e-chg.sep.}}(x_{Li})$ increases leading to a voltage drop (relative to the reference voltage of $CoPO_4$); c) when the intercalation site is in $FePO_4$ far from the interface, $\Delta E_{\text{distortion}}(x_{Li})$, $\Delta E_{\text{e-dipole}}(x_{Li})$ and $\Delta E_{\text{e-chg.sep.}}(x_{Li})$ are zero, and $\Delta E_{\text{e-redox}}(x_{Li})$ is the redox energy difference between Fe and Co, which is about 1.4eV and the Li intercalation voltage is equal to the bulk value of $FePO_4$. A completely analogous argument holds for the other $FePO_4/MPO_4$ interfaces considered in this study. The essential process in the above explanation is the interfacial charge transfer and the associated tradeoff of transition metal redox energy $\Delta E_{\text{e-redox}}$ versus the electrostatic energy cost $\Delta E_{\text{e-chg.sep.}}$ of moving the valence electron away from the Li. This charge transfer process and associated energy terms can quantitatively explain the Li intercalation site voltage profile, as will be shown in **Section 2.2.3**.

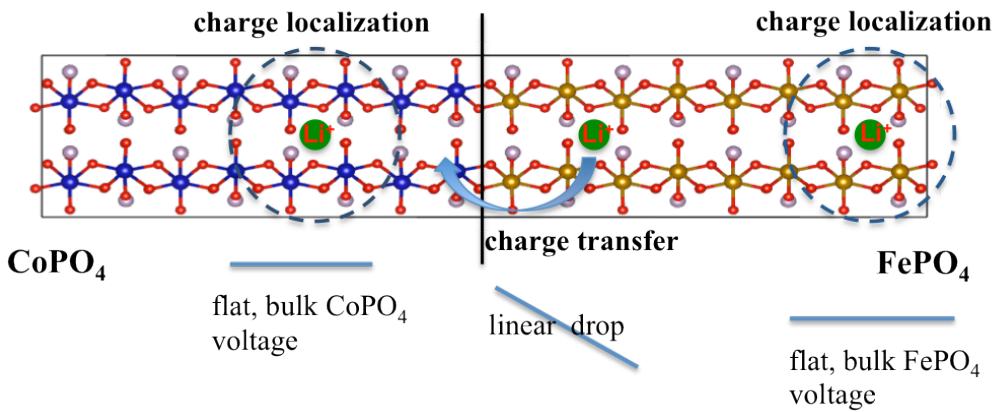

**Figure 2.** Schematic plot showing the charge transfer mechanism in the $CoPO_4/FePO_4$ system with one intercalated Li at different positions (this mechanism analogously holds for all interfaces examined in this study). In the $CoPO_4$ region, the blue dash line around Li indicates the valence electron preferentially localizes on the surrounding Co atoms. When Li is intercalated to the right of the interface, there is charge transfer to the high-voltage region and a corresponding linear voltage drop in the low-voltage side. When the Li is intercalated far from the interface in the low-voltage $FePO_4$ region, the electron localizes on the surrounding Fe atoms and the Li intercalation voltage is the bulk intercalation voltage of $FePO_4$.

In the fully-lithiated limit, the charge transfer mechanism we described above for Li intercalation in the fully-delithiated limit also occurs when a Li atom is deintercalated out of the fully-lithiated cell (beginning of the charging process), and the explanation for the energetics is completely analogous to the process described in the fully delithiated case above. As we mentioned above, the redox energy of Fe is approximately 1.4 eV higher than that of Co. If one Li atom ($Li^+ + e^-$) is removed from the system, and the Li deintercalation position is in the low-voltage region ($FePO_4$), then Fe should be oxidized as a result of removing Li because the Fe is the lower voltage redox. If the Li deintercalation position is in the high-voltage region ($CoPO_4$) and not far from the interface, Fe will still be oxidized but there will be an energy cost due to the electrostatic interaction between the negative charge left by deintercalated $Li^+$ (a Li vacancy, or $V_{Li}$) and the positive charge left by the removed electron (an electron hole). This energy increases when the $V_{Li}$ in the high-voltage region ($CoPO_4$) moves away from the interface, explaining the voltage profile with distance into the high-voltage region. Thus, the physics of this interfacial charge transfer can explain voltage profiles in both the fully-delithiated limit and the fully-lithiated limit. It should be noted here that the hysteresis behavior of the voltage profiles in **Figure 1** would not occur in a real discharging (charging) process because in such a process Li will fill (empty) sites in order of their stability, which is not the order in which we have calculated these site energies. We will consider the more realistic discharge process involving several consecutive Li intercalation steps in **Section 2.3 and 2.4**.

The charge transfer mechanism detailed above can be validated by plotting the charge density difference from our DFT calculations. Here we again use the fully-delithiated limit as an example. In **Figure 3**, we show the charge density difference between the singly-lithiated and the fully-delithiated $FePO_4/CoPO_4$ interface structure.

The Li atom is intercalated at -6.07 Å, 0 Å and 6.07 Å relative to the interface (the interface is placed at 0 Å, and positions in the $CoPO_4$ region are negative). From **Figure 3**, we can see that for the cases where the Li atom is located at the interface and in the $CoPO_4$ region, the Li valence electron localizes on the Co atoms and the charge density difference of Fe atoms is almost zero. In the case where the Li atom is in the $FePO_4$ region near the interface, there is a small charge difference in the $FePO_4$ region (primarily electron gain associated with O atoms in the $FePO_4$ structure), and once again the charge difference of the Fe atoms is nearly zero. This result demonstrates that the electron from the Li is reducing the Co even when Li is located in the $FePO_4$ region, confirming the charge transfer described above.

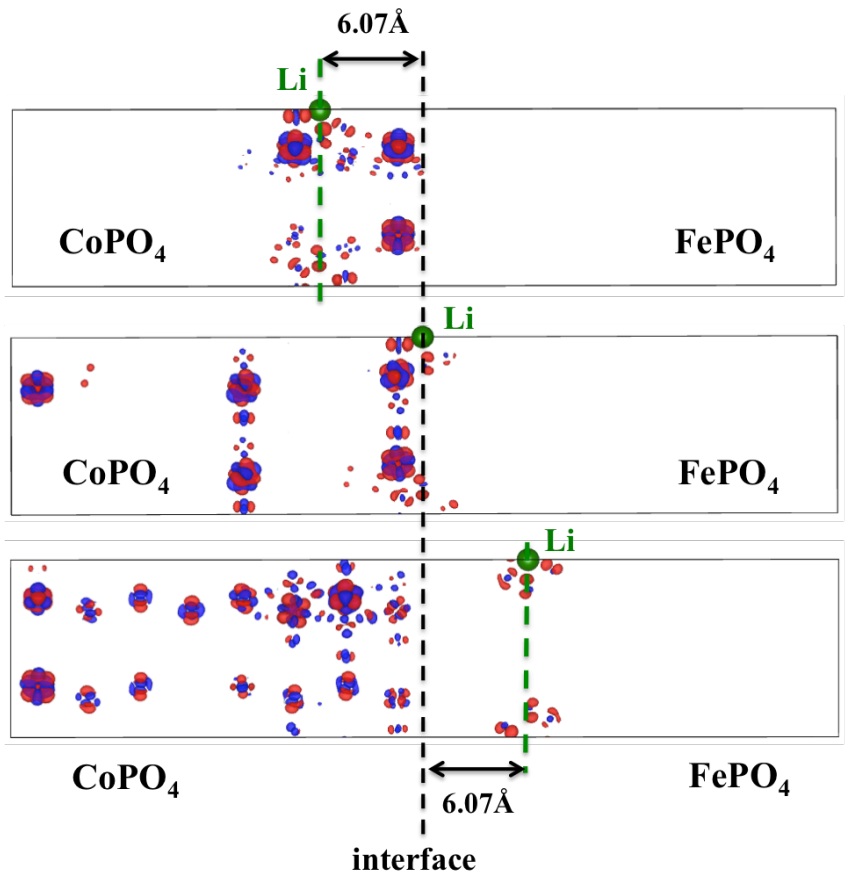

**Figure 3.** Plots of the charge density difference between the fully delithiated $FePO_4/CoPO_4$ and one Li intercalated in $FePO_4$-$CoPO_4$ at different positions (-6.07, 0, 6.07 Å with respect to the interface). The red color region represents positive electron density (electron gain), and the blue color region represents negative electron density (electron loss). The green atom is the intercalated Li.

### 2.2.3 Electrostatic energy calculation

In the previous section, we proposed that redox and electrostatic contributions ($\Delta E_{\text{e-redox}}$ and $\Delta E_{\text{e-chg.sep.}}$) to the Li electron energy explained the voltage behavior shown in **Figure 1** (and that the lattice distortion energy $\Delta E_{\text{distortion}}$ and dipole contribution $\Delta E_{\text{e-dipole}}$ are negligible). To quantitatively validate this assertion we calculated the electrostatic charge-separation energy $\Delta E_{\text{e-chg.sep.}}$ by the Ewald summation method to demonstrate that $\Delta E_{\text{e-chg.sep.}}$ alone is capable of reproducing the voltage profiles in **Figure 1**. We used Bader charge analysis[20] to obtain the charge difference of each atom between the singly-lithiated and delithiated systems, effectively reducing the complicated charge density distribution to a system comprised of point charges located on the atomic sites. The charge differences were used for the Ewald summation to calculate the electrostatic energy of the point charge system. The point charge on each atom was obtained using the formula: Point Charge = Bader charge[Li+Host] − Bader charge[Host] − Bader charge[Li].

The dielectric constants of Fe/Co/Mn/TiPO$_4$ are needed for the Ewald summation. In this work, our designation of "dielectric constant" is meant to imply the zero-frequency (static) dielectric constant of the material in question. We assumed the overall dielectric constant of the interface structure to be the average of the dielectric constants of the two constituent materials in their bulk form. Details on the computational methods used to obtain the dielectric constants can be found in **Section 5**. The calculated dielectric constants of FePO$_4$, CoPO$_4$, MnPO$_4$, TiPO$_4$ are 4.2, 5.8, 5.1, 3.3 respectively. The dielectric constant of FePO$_4$ calculated by a previous computational work is 3.8[21], in good agreement with our present result.

A comparison of the electrostatic energy $\Delta E_{\text{e-chg.sep.}}$ results with the DFT results in the fully-delithiated limit are shown in **Figure 4**. Here, we transfer the unit of $\Delta E_{\text{e-chg.sep.}}$ from eV to V by the unit change $\Delta E_{\text{e-chg.sep.}}/e$ to show clearly the voltage variation across the interface caused by this electrostatic energy term. In each FePO$_4$/MPO$_4$ case of **Figure 4**, the $\Delta E_{\text{e-chg.sep.}}/e$ profile is shifted by a constant to align to the voltage plateau of the material with higher Li intercalation voltage. This shift is necessary because the electrostatic energy $\Delta E_{\text{e-chg.sep.}}/e$ of the charge difference itself cannot give the correct quantitative voltage value, therefore we just compare the changes in the electrostatic and DFT voltage curves to show the impact of the electrostatic charge-separation energy due to the charge transfer on the voltage profile. The voltage profiles predicted by the electrostatic energy calculation match very well with the DFT results at the high-voltage plateau and linear voltage drop region. This result further validates the charge transfer mechanism and proves that the electrostatic energy $\Delta E_{\text{e-chg.sep.}}$ cost of the charge transfer across the interface leads to the approximately linear voltage drop in the low-voltage region. The discrepancy between the electrostatic energy $\Delta E_{\text{e-chg.sep.}}/e$ results and DFT results around the low-voltage plateau is due to a portion of the valence electron charge of Li localizing on the transition metal atoms of the low-voltage region. The charge transfer to the high-voltage region tends to decrease and eventually stop as the Li intercalation site proceeds farther from the interface into the low-voltage region. As the redox energies $\Delta E_{\text{e-redox}}$ of different transition metal atoms are different, the electrostatic energy calculation cannot reproduce this redox energy difference, resulting in the quantitative differences between the electrostatic and DFT results around the low-voltage plateau.

The above single Li intercalation results and the proposed charge transfer mechanism not only aid in understanding the physics governing the Li intercalation voltage profile at the interface, but also have direct implications for redox pseudocapacitors. We will discuss the implications of the single Li intercalation results in **Section 3**.

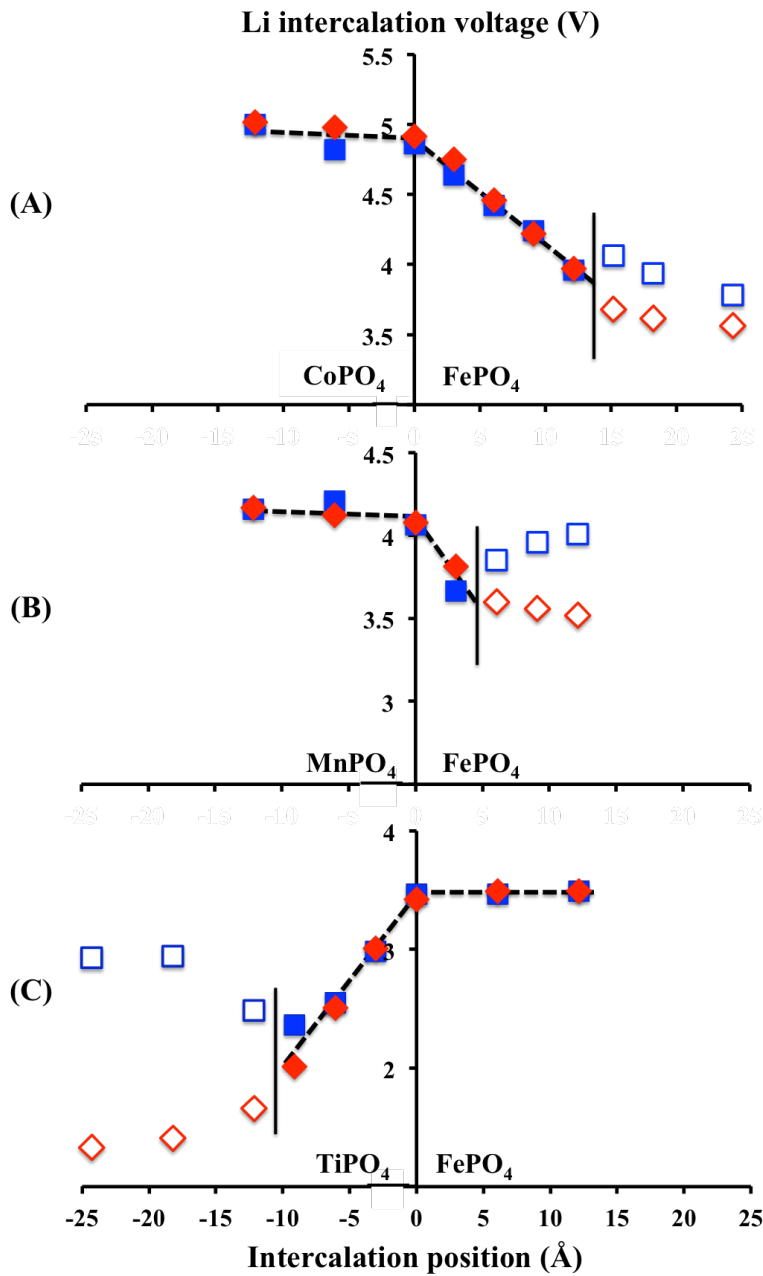

**Figure 4.** Comparison of the electrostatic charge-separation energy $\Delta E_{\text{e-chg.sep.}}/e$ results with the DFT results of the Li intercalation voltage profiles across the MPO$_4$/FePO$_4$ interfaces. (a) CoPO$_4$/FePO$_4$; (b) MnPO$_4$/FePO$_4$; (c) TiPO$_4$/FePO$_4$. In all cases, the blue (square) symbols denote electrostatic energies obtained from the Ewald method and the red (diamond) symbols denote DFT data. The electrostatic energy profile is shifted to align to the higher Li intercalation voltage plateau. The region where the DFT and electrostatic data diverge is represented by the hollow data points. In (a)-(c), the black dashed lines are included as guides to the eye.

## 2.3 Screening effect of the interfacial dipole modified by the intercalated Li

In the above calculations of Li intercalation site voltage profiles, we only considered the process of intercalating/deintercalating a single isolated Li atom, illustrating the general energetics of Li in the system. However, in real discharging/charging processes, Li atoms go into/out of the cathode material one by one in the intercalation/deintercalation order of high→low/low→high voltage sites. Because of this order there are always multiple Li atoms that have already been intercalated/deintercalated into/out of the cathode. Therefore, to investigate this more realistic process we examined the influence of multiple Li atoms present in the interface system on the intercalation voltage profile. The discharging process (Li intercalation) is used as the representative case in the following discussion.

In **Figure 5** we explicitly calculate the intercalation voltages of the sites in the low-voltage material ($FePO_4$) near the interface by filling the sites in their order of stability. Due to the finite size periodic cell we must fill an entire plane of Li intercalation sites for each Li atom we add to the cell, and each layer has only two sites in the cell. Therefore, when filling in order of stability we fill four sites in the first two layers. From **Figure 5**, the voltage enhancement effect exists only in the first layer closest to the interface. The voltage drops quickly to the voltage plateau of $FePO_4$ in the second layer. This reduction in the range of the voltage enhancement in the layer-by-layer intercalation is due to screening from the Li already in the system impacting the energetics of new Li being added. In **Section 2** of the **SI** we demonstrate that this screening alters the voltage by increasing the interfacial dipole across the interface, i.e., increasing the $\Delta E_{\text{e-dipole}}$ term in **Equation 1**. More specifically, the interfacial dipole is enhanced when Li atoms are intercalated in the low-voltage region near the interface, and when charge transfer of the intercalated Li valence electron from the low-voltage material to the high-voltage material across the interface occurs. Once the interfacial dipole is enhanced in the system from a Li valence electron transfer, as more Li is intercalated in the low-voltage region it will cost progressively more electrostatic energy to overcome the interfacial dipole and transfer the additional Li valence electrons from the low- to the high-voltage region (equivalent to an increase of the $\Delta E_{\text{e-dipole}}$ energy). The effect of this interfacial dipole modification makes it more difficult for the charge transfer from the intercalated Li in the low-voltage region to utilize the transition metal redox states in the high-voltage region. Thus, the voltage enhancement effect in the low-voltage region near the interface is suppressed compared with the results shown in **Figure 1**. This interfacial dipole screening effect will always occur for a Li intercalation process that is relatively close to equilibrium. However, the magnitude of screening from the dipole could potentially be greatly reduced during very fast intercalation, where the Li sites do not fill in order of stability but instead have occupations controlled by kinetic factors.

In considering a realistic intercalation process that is close to equilibrium it is also important to realize that the high-voltage material will intercalate first. For the low-voltage material to experience any enhancement under such conditions there must be active redox states available in the fully lithiated high-voltage material. We consider such an interface in **Section 2.4** using $LiNiO_2$, and also explore how the previous results are altered by changing from an olivine structure to a layered structure.

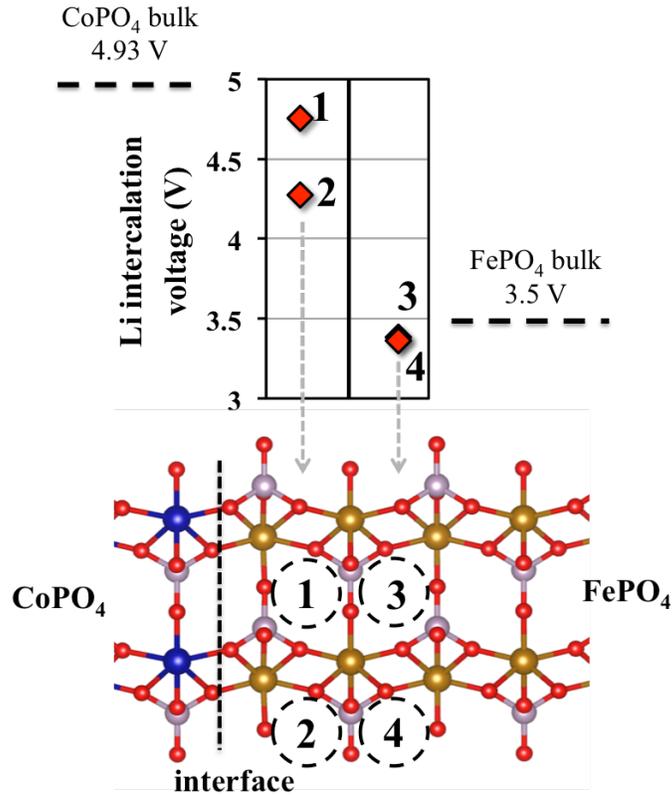

**Figure 5.** Li intercalation site energies in $CoPO_4$-$FePO_4$ where the $FePO_4$ sites are gradually filled up layer by layer. The x-axis represents Li intercalation site which shows the order of $FePO_4$ intercalation sites being occupied. For example, atom 1 intercalates into the position

shown in the interfacial structure plot. After atom 1 is inserted, additional Li are sequentially intercalated into their corresponding sites in the structure.

## 2.4 Utilization of multiple-redox transition metal atoms with a LiNiO$_2$/TiO$_2$ interface

As we discussed in **Section 2.3**, in a real discharge process Li atoms will first occupy the high voltage intercalation sites. Using FePO$_4$/CoPO$_4$ as an example, CoPO$_4$ will be completely lithiated first, and the redox state change of Co will be Co$^{3+}$→Co$^{2+}$. If more Li is subsequently intercalated in the FePO$_4$ region, the utilization of Co redox states is not possible because all Co atoms are reduced to Co$^{2+}$ in fully-lithiated LiCoPO$_4$, and it is energetically unfavorable to further reduce Co$^{2+}$. Thus, the availability of transition metal redox states in the high-voltage material creates an intrinsic limitation of the voltage enhancement effect in the low-voltage material.

The limitation just described can be overcome by creating an interface containing materials with a transition metal element that has extra redox states available to be reduced when the material is fully lithiated. This can be done without sacrificing capacity but using transition metals with multiple redox states allowing more than one electron of redox all active at high voltage. It has been previously reported that Ni in the layered-structured, e.g. Li$_x$(Ni, Mn, Co)O$_2$ (NMC cathode material) progresses through the redox states Ni$^{4+}$→Ni$^{2+}$ during the discharging process at voltages above 3.6 V [22]. Therefore, LiNiO$_2$ meets our requirement of having remaining high-voltage reducible atoms even when fully lithiated (i.e., even when all Li sites in the structure are occupied). We then constructed an interface, choosing TiO$_2$ as the low-voltage layered-structure material because its Li intercalation voltage (calculated to be 2.4 V for the first Li intercalation in TiO$_2$, i.e. TiO$_2$→Li$_{0.014}$TiO$_2$) is significantly below NiO$_2$ (calculated to be 4.3 V for the first Li intercalation in NiO$_2$, i.e. NiO$_2$→Li$_{0.014}$NiO$_2$) and it possesses the same layered-structure as NiO$_2$ so that complicated lattice matching interface effects can be avoided.

In our exploration of the Li intercalation energetics in this new model interface system, the NiO$_2$ region is already completely lithiated (LiNiO$_2$) and Li will be intercalated in TiO$_2$ at different positions. The Li intercalation site voltage profile in TiO$_2$ is shown in **Figure 6**. It is clear the voltage enhancement is quite dramatic in the low-voltage material TiO$_2$. Li intercalation voltage in the TiO$_2$ side is very close to our calculated bulk voltage of $NiO_2$ (4.3 V) and significantly higher than our calculated bulk voltage of $TiO_2$ (2.4 V). Compared with the single-Li intercalation voltage profile in the olivine-structured case, the voltage-drop rate in $TiO_2$ is very small. Several $LiNiO_2/TiO_2$ calculations were therefore performed with longer supercells than shown in **Figure 6** to ascertain the behavior of Li intercalation site voltage very far from the interface, and the results are shown in **Section 3** of the **SI**. As expected, the voltage decreases when the intercalation position is far enough away from the interface (>20 Å), and we estimate that the Li intercalation site voltage would decay to the $TiO_2$ bulk value at ~9 nm (**SI Section 3, Figure S4)**. We found that of the terms in Equation 1 a combination of electrostatic energy cost $\Delta E_{\text{e-chg.sep.}}$ induced by charge transfer (discussed in **Section 2.2**) and a strain energy contribution $\Delta E_{\text{distortion}}$ caused by Ti atom distortion near the intercalated Li explains the Li intercalation voltage behavior shown in **Figure 6**. A more detailed analysis of the single-Li intercalation voltage profile shown in **Figure 6** and explanations for the difference of voltage profile behavior between the olivine-structured interface and the layered-structured interface are provided in **Section 3** of the **SI**.

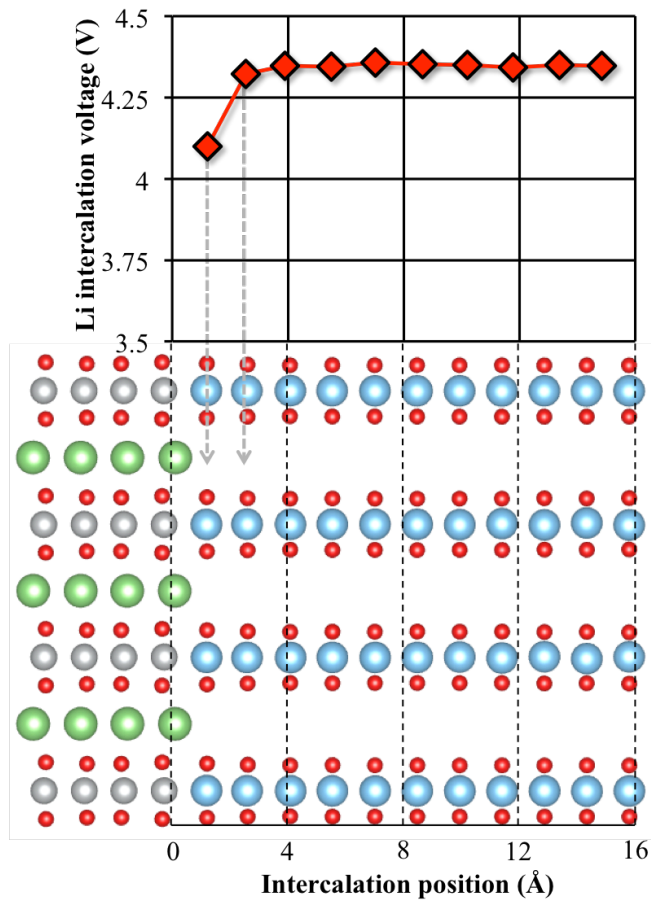

**Figure 6.** Single Li intercalation voltage profile in the LiNiO$_2$/TiO$_2$ interface. LiNiO$_2$ is already fully lithiated, and Li atoms are intercalated in TiO$_2$ at different positions. The calculated Li intercalation voltage at different TiO$_2$ sites are represented by red diamond points. Li intercalation voltage in TiO$_2$ is very close to our calculated bulk voltage of NiO$_2$ (4.3 V) and significantly higher than our calculated bulk voltage of TiO$_2$ (2.4 V).

As in **Section 2.3** for the olivine structures, we also consider a realistic intercalation process where we insert Li into intercalation sites in TiO$_2$ layer by layer, filling the sites in approximate order of their stability. The calculated layer-by-layer Li intercalation voltage profile in the LiNiO$_2$/TiO$_2$ interface is shown in **Figure 7**. As with the olivine case, the intercalated Li acts to screen the interfacial voltage enhancement, and from **Figure 7** the interfacial voltage enhancement is shown to occur in just the first few atomic layers of TiO$_2$, or over ~0.5 nm. The voltage drop rate in TiO$_2$ in this case is much faster than the voltage drop rate in the Li intercalation site voltage profiles shown in **Figure 6**. This increase in the voltage drop rate with layer-by-layer Li intercalation is again due to the increasing interfacial dipole $\Delta E_{\text{e-dipole}}$ energy, as discussed for the olivine case in **Section 2.3**.

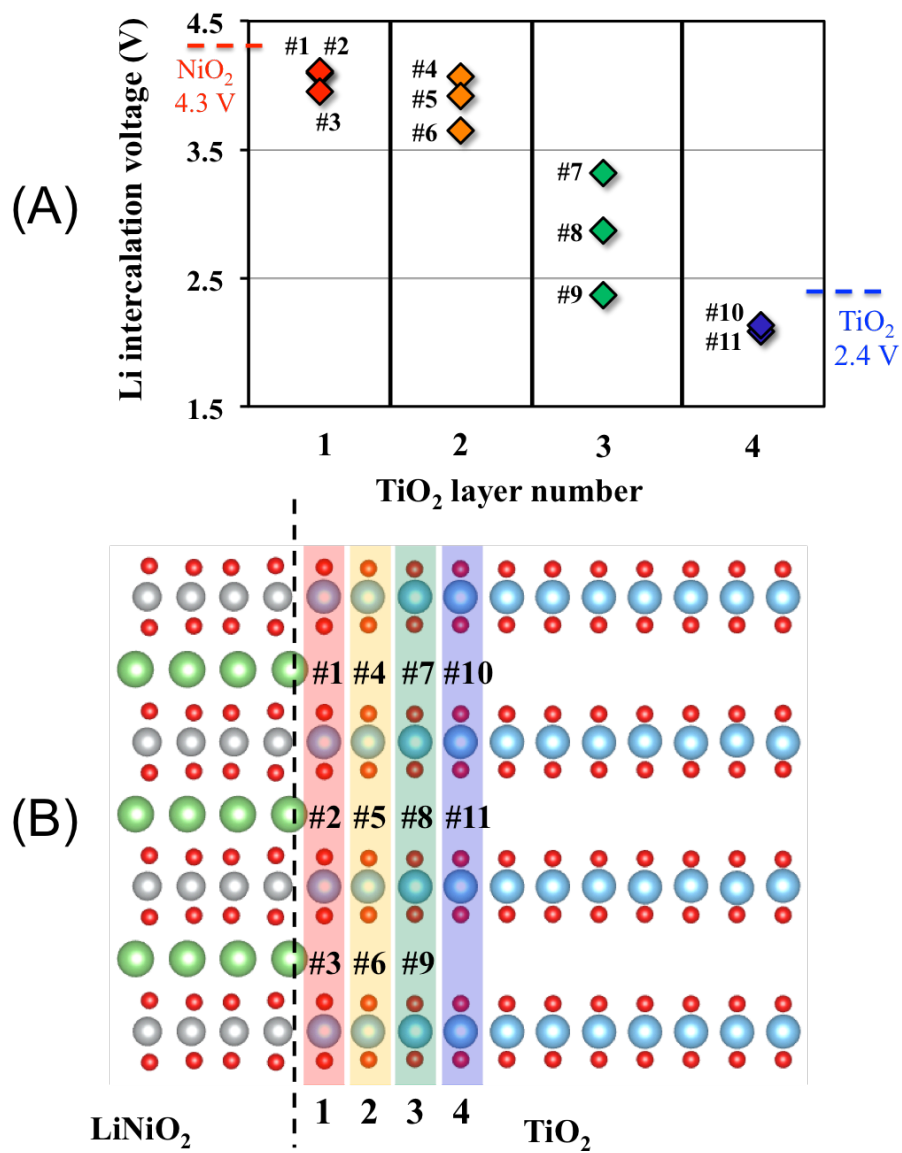

**Figure 7.** Li intercalation site energies in LiNiO$_2$/TiO$_2$ system where Li sites in TiO$_2$ are filled layer-by-layer. In (A), the red, orange, green and blue diamonds represent the voltage values in the first, second, third and fourth layer respectively. In (B), the LiNiO$_2$/TiO$_2$ structure is shown, where the colored bands and layer labels correspond to the Li insertion voltages and layer positions in (A). The order of Li intercalation is the same as shown in **Figure 5**.

# 3 Implications of the interfacial voltage enhancement

## 3.1 A possible Li-ion battery cathode heterostructure design

An important implication of the nanoscale voltage enhancement effect detailed in this work is that the Li intercalation voltage in the low-voltage material can be increased by the high-voltage material near the interface. As shown in **Figure 7**, the voltage enhancement region can extend several intercalation layers (~0.5nm). To utilize this effect one can create a superlattice structure with the intergrowth of high- and low-voltage materials, and the low-voltage material would be effectively shifted to high-voltage by the interfaces present on both sides of the low-voltage material. This shift can allow optimization of the low-voltage material for other properties, as its voltage is no longer a constraint. For example, **Figure 8** shows a schematic of a multivalent cathode heterostructure with more than one redox state per Li (e.g. $Li_xNiO_2$) (red region) and a material optimized for fast Li transport (blue region). The "red" material effectively activates the "blue" material through the interfacial voltage effect, so there is little loss in energy density compared to the case of having a material comprised of all high-voltage ("red") material. However, the "blue" material can provide fast diffusion paths, enabling rapid charge and discharge. Some possible specific pairs of the "red" and "blue" materials could be $Li_xNiO_2$ as the high-voltage "red" material and a solid electrolyte material (e.g. $Li_{10}GeP_2S_{12}$ or $Li_7P_3S_{11}$) as the low-voltage "blue" material. The Li diffusivity in the layered-structured $Li_xMO_2$ material is relatively low, in the range of $10^{-8}$ to $10^{-13}$ $cm^2/s$.[23-26] In the solid electrolyte "blue" materials $Li_{10}GeP_2S_{12}$ and $Li_7P_3S_{11}$, the diffusivity is faster, in the range of about $10^{-6}$-$10^{-7}$ $cm^2/s$, which is ≥2 orders of magnitude faster than that in the "red" $Li_xMO_2$. This heterostructured cathode material could be intercalated primarily through the fast "blue" material, requiring Li migration through only a few nanometers in the "red" high-voltage material. The presence of the low-voltage "blue" solid electrolyte material can therefore enhance the Li transport properties of the overall heterostructure compared to a pure "red" material, while the high-voltage "red" $Li_xMO_2$ material ensures a high energy density in the "blue" region due to the voltage enhancement effect. This heterostructure design can provide us the possibility of making the cathode from solid bulk materials instead of micron-scale particles, which would decrease the contact area between cathode material and organic electrolyte solution significantly, leading to less electrolyte decomposition during charging/discharging cycles. It should be noted here that we have not considered the complicated issues of the growth of this heterostructure design into a functional superlattice material and the issues of the lattice mismatch at the interface. The lattice mismatch may cause stability issues in the superlattice structure and may also change the atomistic structure of the low-voltage ("blue") material, perhaps altering its Li transport properties. Therefore, further exploration for ways to create such a heterostructured material, including finding good candidate materials for the "red" and "blue" pair, are still needed in the future.

The above idea of a heterostructured cathode material may be particularly useful for multivalent battery cathodes (e.g. Mg-ion battery and Al-ion battery), where it is difficult to find materials with both good cation diffusion and high voltage. For example, $V_2O_5$ is a relatively high voltage cathode material for a Mg-ion battery, with an open circuit voltage 2.66V vs. Mg/Mg$^{2+}$[27], but the diffusion of Mg ions in $V_2O_5$ is impractically slow for many applications[28]. On the other hand, molybdenum sulfide materials, like chevrel phase $Mo_6(S/Se)_8$[29] and layered-structured $MoS_2$[27], have relatively fast Mg ion transport, but have voltages that are significantly lower than that of $V_2O_5$ ($Mo_6S_8$ ~1.2V, $MoS_2$~1.5V). A heterostructure cathode material like that proposed in **Figure 8** might consist of $V_2O_5$ as the "red" high-voltage material and a fast diffusing sulfide as the "blue" material. For the Al-ion battery, one recently published work showed an ultrafast rechargeable Al battery cell[30], where the battery can be rapidly charged/discharged at a current density up to 4000 mA/g (~60C rate) for more than 7000 cycles with almost no loss in capacity. The cathode material used in this cell is graphite, which has a relatively low voltage (2V vs. Al/Al$^{3+}$). However, based on our heterostructured cathode design, graphite could be a good candidate for the "blue" low-voltage material that gives very fast Al ion transport. For the "red" high-voltage material, further work is needed, but researchers could focus on discovering materials with high Al intercalation voltage, without needing particularly good Al transport.

The above heterostructure design could also potentially be adapted to improve the cyclic stability of the high voltage/capacity cathode materials' structures in Li batteries (e.g. Li-rich NMC cathode) by using a low-voltage material as a protection layer that stabilizes the high-voltage material during charging/discharging cycles. Another use of such a heterostructure might be if the low-voltage material (the blue material in **Figure 8**) is very inexpensive, the cost of the cathode material might be lowered by this heterostructure design while keeping the same energy density. This heterostructure design is also analogous to the material intergrowth existing in Li-excess cathode $x$Li$_2$MnO$_3 \cdot$(1− $x$)LiMO$_2$ (M=Ni, Co, Mn, Fe, etc.). Previous experimental work[31] showed that the thickness of the intergrowth layer in Li-excess cathode material is about 1-3 nm (shown in STEM images), which is comparable to our predicted effective length scale of voltage enhancement (~1 nm in the context of material intergrowth due to double interfaces on both sides of the layer). Our findings detailed in this work imply that the voltage activation effect may play a critical role in understanding the Li intercalation energy in Li-excess cathodes because a significant portion of the relatively low-voltage material (LiMO$_2$ ~4V [32]) might be activated by the high-voltage material (Li$_2$MnO$_3$ ~4.5V [32]).

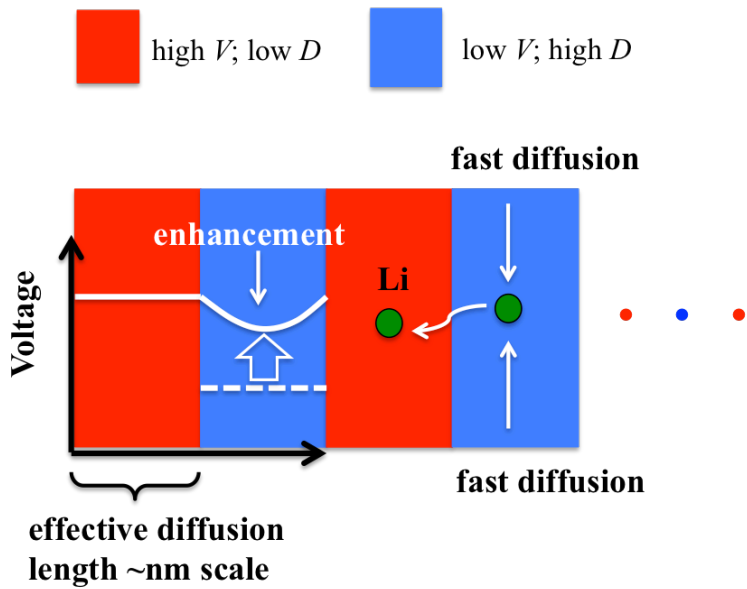

**Figure 8.** Schematic plot of a cathode superlattice structure (intergrowth of two materials). The solid white curve shown in the left two layers represents the intercalation voltage profile. The dashed white line denotes the bulk intercalation voltage of the blue, low-voltage material. The intercalation voltage of the low-voltage material is enhanced due to the interfaces with the high-voltage material on both sides. The diffusion path for a Li atom (green circle) has two parts: Li first diffuses from the boundary of the cathode particle to the blue (low-voltage) material and then

diffuses from the blue (low-voltage) material to the red (high-voltage) material. The effective diffusion length in the slower high-voltage material is then just approximately equal to the thickness of the high-voltage layer.

At the end of this section, we want to briefly discuss the voltage measurement of this high-/low-voltage heterostructure system at the initial stage of the Li intercalation process (discharging). Under the assumption of quasi-equilibrium discharging condition (very slow rate), although the Li atoms first diffuse through the low-voltage material region (fast Li transport), we will always measure the high voltage value at the initial discharging stage no matter whether there is significant voltage enhancement effect in the low-voltage material, this is mainly because of the entropic effect that always drives the Li into the high-voltage material to make the system get into the lowest-free-energy state (thermodynamic equilibrium state) under the quasi-equilibrium discharging condition. So it is the voltage value of the high-voltage site being represented by the voltage measurement at the initial stage of the discharging process.

## 3.2   Implications for redox pseudocapacitors

The predicted Li intercalation site voltages also have important implications for the thickness of coating layers that might be used in redox pseudocapacitors. The redox pseudocapacitor stores electrochemical energy by adsorbing ions onto or near the surface of the electrode (usually a multivalent transition metal oxide), resulting in a faradaic charge transfer and electrochemical reduction of the transition metal atoms in the electrode.[19] Because there is no mass transport of ions into the electrode in redox pseudocapacitors, these devices usually have much higher charging/discharging rates than batteries. Pseudocapacitors generally use a solid electrode to contribute the faradaic capacitance and a liquid electrolyte for rapidly transporting the ions to/from the electrode surface. One common issue for the redox pseudocapacitor is electrode degradation, which limits the lifetime of the device. One example is the degradation of $MnO_2$ (one of the most widely used materials as the redox pseudocapacitor electrode) under acidic and near neutral pH conditions[33]. This degradation can be mitigated by appropriate electrode engineering methods. For example, one previous experimental work demonstrated that the capacity fade can be significantly reduced by increasing the binder (electrochemically inactive materials) content in the composite electrode[34]. It may also be possible to deposit a protective coating layer (another electrochemically inactive material) on the electrode surface to prevent the electrode degradation, which is similar to what is done for Li-ion battery cathodes [35].

An important issue to understand is the influence of any inactive protective layer on the faradaic charge transfer, and our single Li intercalation results provide useful insights into this issue. In the following discussion, we assume there is a conformal inactive layer around the active electrode, which forms an interface between a high- and low-voltage material, similar to the interfaces studied in this work. As there is no intercalation process of the ions into the composite electrode (active material), the interfacial dipole screening effect due to the presence of multiple Li in the inactive material can be neglected. Therefore, our single Li intercalation model in the olivine-structured (**Section 2.2**) and the layered-structured material (**Section 2.4**) can qualitatively show the effective length of faradaic charge transfer across the inactive material. From our previous calculations, the olivine-structured material ($FePO_4/MPO_4$) has an effective length for the faradaic charge transfer of about 1-1.5 nm. In the layered-structured $LiNiO_2$-$TiO_2$ case this length is about 9 nm. These results therefore provide guidance for the optimal thickness of the protective layer on the redox pseudocapacitor electrodes, and implies that the faradaic charge transfer (charge separation) across a nanometer-scale inactive material is possible.

# 4 Conclusions:

Overall, this study shows that the general behavior of the Li intercalation voltage across cathode interfaces between two materials with different intercalation voltages is: (a) a plateau in the high-voltage material, equal to the bulk Li intercalation voltage of the high-voltage material, (b) a voltage drop in the low-voltage material over a nanometer-scale distance from the interface, and (c) a plateau when the intercalation site is far enough away from the interface in the low-voltage material, where the value approaches the bulk intercalation voltage of the low-voltage material. The nanoscale voltage enhancement effect in the near-interface region in the low-voltage material occurs because electrons are transferred across the interface to the high-voltage material. We have shown the voltage changes at the interface are due to a combination of strain, redox, dipole, and electrostatic effects. The interfacial voltage enhancement might be utilized to enhance performance of Li-ion battery cathodes by enabling use of nanoscale layers of materials that would otherwise have a voltage that is too low to be practical. Examples might include heterostructures with fast Li diffusers and protective coatings. A possible implication of this effect is that one may design a superlattice cathode structure in which the advantageous properties of the low-voltage material (e.g., high Li conductivity, electronic conductivity, or thermodynamic stability) may be utilized without significant voltage or energy density loss due to the voltage enhancement from the high-voltage materials. We also note that our single Li intercalation calculations and the charge-transfer electrostatic energy calculation are relevant to redox pseudocapacitors and the associated faradaic charge transfer mechanism. Specifically, pseudocapacitors sometimes use inactive materials as protective layers on the electrodes to reduce degradation. Our results suggest the thickness limits of inactive protective coating materials above which the faradaic charge transfer will not occur effectively. The nanoscale voltage enhancement effect and understanding of the governing physics of Li intercalation energetics at interfaces provides insight in the future design of high-performance electrochemical energy storage systems.

# 5  Computational methods:

All calculations were performed with density functional theory (DFT) using the in the Vienna Ab Initio Simulation Package (VASP) [36, 37] code. The projector augmented wave method (PAW)[38] was used for the effective potential for all atoms. The PAW potentials used in these calculations have valence electron configurations of $2s^22p^4$ for O, $3s^23p^3$ for P, $1s^22s^1$ for Li, $3p^63d^74s^1$ for Fe, $3d^84s^1$ for Co, $3p^63d^64s^1$ for Mn, $3p^63d^34s^1$ for Ti and $3p^63d^94s^1$ for Ni. The generalized gradient approximation exchange-correlation functional of PW-91[39] was used with the Hubbard $U$ correction (GGA+$U$)[40] applied to the transition metal atoms. The effective $U$ values for different transition metal atoms are provided in **Table S2** of **SI Section 1.2**. The U values are obtained from the previous theoretical work[41] where the U values for the different transition metal elements were chosen to match with the experimental Li intercalation voltages in different transition metal oxide cathode materials (olivine, layered-structured materials). A 500 eV plane wave energy cutoff was large enough for the calculated voltage value converged within 10 meV/Li. A Γ-centered k-point mesh was used for the Brillouin zone sampling for Li intercalation site energy calculations to obtain Li intercalation voltages converged within 30 mV/Li. The k-point mesh for the bcc Li metal (conventional lattice constant is 3.45 Å; 2 Li atoms per supercell) was 15x15x15 with an energy convergence better than 1 meV/atom.

In this work, we represent the Li intercalation energy at different insertion positions in terms of voltage. The Li intercalation voltage can be approximated using DFT calculations as [42]:

$$V = \frac{E(\text{Li+Host}) - E(\text{Host}) - E(\text{Li metal})}{e} \quad (2)$$

where $E(\text{Li+Host})$ is the energy of an olivine-structured or layered-structured supercell after a Li is intercalated, $E(\text{Host})$ is the energy of the supercell before the Li is intercalated, $E(\text{Li metal})$ is the energy of Li metal. The expression in the numerator of **Equation 2** is the formation energy of one intercalated Li and dividing this quantity by the fundamental electron charge $e$ (including the negative sign) yields the intercalation voltage. In this way, the intercalation voltage of a Li atom is just the negative of its formation energy. In our calculations of the Li intercalation site voltage, we ignore the finite temperature and configurational entropy $S_{config}$ terms throughout as these terms are approximately $k_BT=26$ meV at room temperature (300K) and would have no significant impact on the results, which are typically on the eV scale, as shown in Figure 1 and Figure 7.

For the olivine-structure interface, the simulated olivine unit cell has orthorhombic symmetry and space group *Pnma* (#62). We duplicated the olivine unit cell along the <010> direction. The interface is perpendicular to the <010> direction of the olivine unit cell, which is the fastest Li diffusion direction. Similar approaches were used for the creation of the layered-structured interface. The symmetry of the layered-structured unit cell is rhombohedral and corresponds to the space group $R\bar{3}m$ (#166). The interface is perpendicular to the Li-containing transition metal oxide layer. All interface supercells were long enough to ensure the intercalated Li atom was only affected by a single FePO$_4$-MPO$_4$ interface or a single LiNiO$_2$/TiO$_2$ interface. Refer to the **SI Section 1.1** for the details of interface cell structures.

To calculate the dielectric constants for each pristine olivine MPO$_4$ (M=Fe, Co, Ti) material, density functional perturbation theory (DFPT)[43] implemented in the Vienna Ab Initio Simulation Package (VASP) code was used. The convergence criterion of the structural relaxation for the DFPT calculation was 10$^{-6}$ eV/cell to ensure the atomic forces were less than 0.001 eV/ Å.

## ACKNOWLEDGEMENTS

The authors gratefully acknowledge funding from the Dow Chemical Company and helpful conversations with Mark Dreibelbis, Brian Goodfellow, Mahesh Mahanthappa, and Robert Hamers. Computations in this work benefitted from the use of the Extreme Science and Engineering Discovery Environment (XSEDE), which is supported by National Science Foundation grant number OCI-1053575.

**Table of contents**

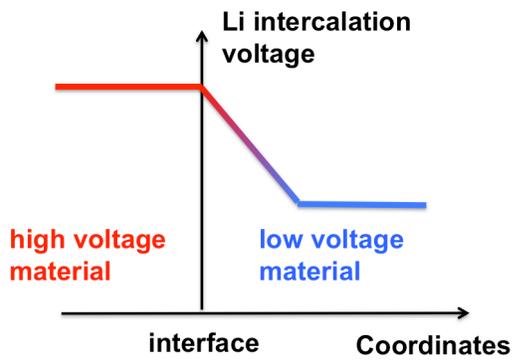 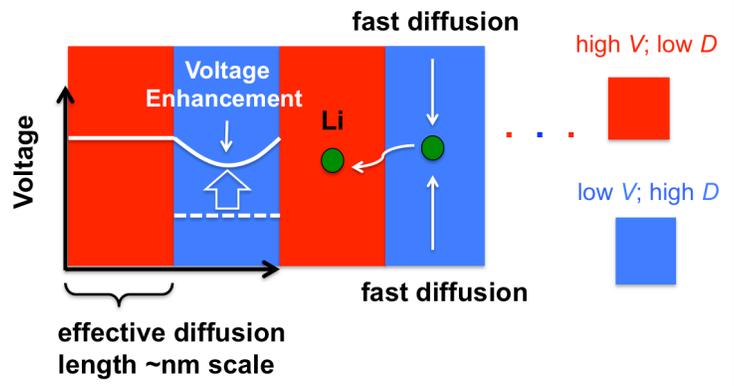

## Supporting Information

Details of the interface structure setup, Hubbard U values, the influence of the interfacial dipole on the electrostatic potential and the supplemental explanations for the results in **Figure 6** (main text).